
\documentclass[twocolumn]{aastex63}
\usepackage{epsfig}                     % For eps figures, old commands
\usepackage{graphicx}                    % For eps figures, newer & more powerfull
\usepackage{courier}     
\usepackage{natbib}                 % Change the \texttt command to courier style
\usepackage{amssymb}                    % useful mathematical symbols
\usepackage{color}                       % For color text: \color command
\usepackage{hyperref}
\usepackage[space]{grffile}

\graphicspath{{./}{figures/}}
\begin{document}

%\title{High speed streams from two origins triggering geomagnetic activity on the solar %cycle 23}
%\title{On the geomagnetic storms and  the Russell-McPherron effect at the Lagrange Point %L1 and the magnetosphere}
\title{Compton-Getting effect due to terrestrial orbital motion observed on cosmic ray flow from Mexico-city Neutron Monitor}

\author{C.E. Navia}
\affiliation{Instituto de F\'{i}sica, Universidade Federal Fluminense, 24210-346, Niter\'{o}i, RJ, Brazil }
\correspondingauthor{C.E. Navia}
\email{carlos\_navia@id.uff.br}

\author{M.N. de Oliveira}
\affiliation{Instituto de F\'{i}sica, Universidade Federal Fluminense, 24210-346, Niter\'{o}i, RJ, Brazil }

%\author{M.C.A. Vieira}
%\affiliation{Instituto de F\'{i}sica, Universidade Federal Fluminense, 24210-346, Niter\'{o}i, RJ, %Brazil }

\author{A.A. Nepomuceno}
\affiliation{Departamento de Ci\^encias da Natureza, Universidade Federal Fluminense, 28890-000, Rio das Ostras, RJ, Brazil}

\begin{abstract}
We look for a diurnal anisotropy in the cosmic ray flow,
 using the Mexico-City Neutron Monitor (NM) detector, due to  the Earth's orbital motion and predicted by Compton-Getting (C-G) in 1935, as
a first-order relativistic effect. The Mexico-City NM's geographic latitude is not very high ($19.33^{\circ}$N), and it has a high cutoff geomagnetic rigidity (8.2 GV) and mountain altitude (2274 m asl) favoring the observation of the C-G effect. 
Furthermore, during the solar cycle minima, the galactic cosmic ray flux is maxima, and the solar magnetic field gets weakened, with a dipolar pattern. Its influence on cosmic rays reaching Earth is the smallest.
Analysis of the combined counting rate during two solar minima,  2008 and 2019, from Mexico-city NM's data yields the C-G effect with an amplitude variation of (0.043$\pm$ 0.019)\%, and phase of (6.15$\pm$ 1.71) LT. The expected amplitude variation is 0.044\%, and the phase of 6.00  LT. 
\end{abstract}


\section{Introduction} 
\label{sec1}

The observed particle distributions in two frames of reference in relative motion are different. For instance, if the particle distribution is isotropic in a given reference frame, it must have an anisotropy in a reference frame in relative motion to the previous one in the direction of movement. That effect is known as the Compton-Getting (C-G) effect \citep{comp35}. 
Thus, considering that galactic cosmic rays have an almost isotropic distribution into the heliosphere (as is expected during the solar cycle minima), it is expected an anisotropy in the daily distribution of cosmic ray intensity at Earth due to the Earth's orbital motion around the Sun. The cosmic ray intensity should be higher, coming from the direction the Earth is moving, i.e., around 06:00 LT \citep{cutl86}.

However, as Earth's orbital velocity is relatively small, having an average value of 29.78 km/s, observation of the C-G effect requires high-energy cosmic particles to minimize distortions due to interplanetary magnetic field, the solar wind, and the Earth's magnetic field.

The Earth's orbital motion is measured using underground muon flux because underground muons come from galactic cosmic rays with high geomagnetic rigidity cutoff  (above 1000 GV). Indeed, it is reporting an anisotropy due to the C-G effect, with an amplitude of about 0.025\%. The parent particles were galactic cosmic rays with a stiffness of around 1.5 TeV/c during an observation period of 5.4 years \citep{cutl86}.

\begin{figure*}[]
\vspace*{-0.0cm}
\hspace*{+0.0cm}
\centering
\includegraphics[clip,width=0.8
\textwidth,height=0.4\textheight,angle=0.] {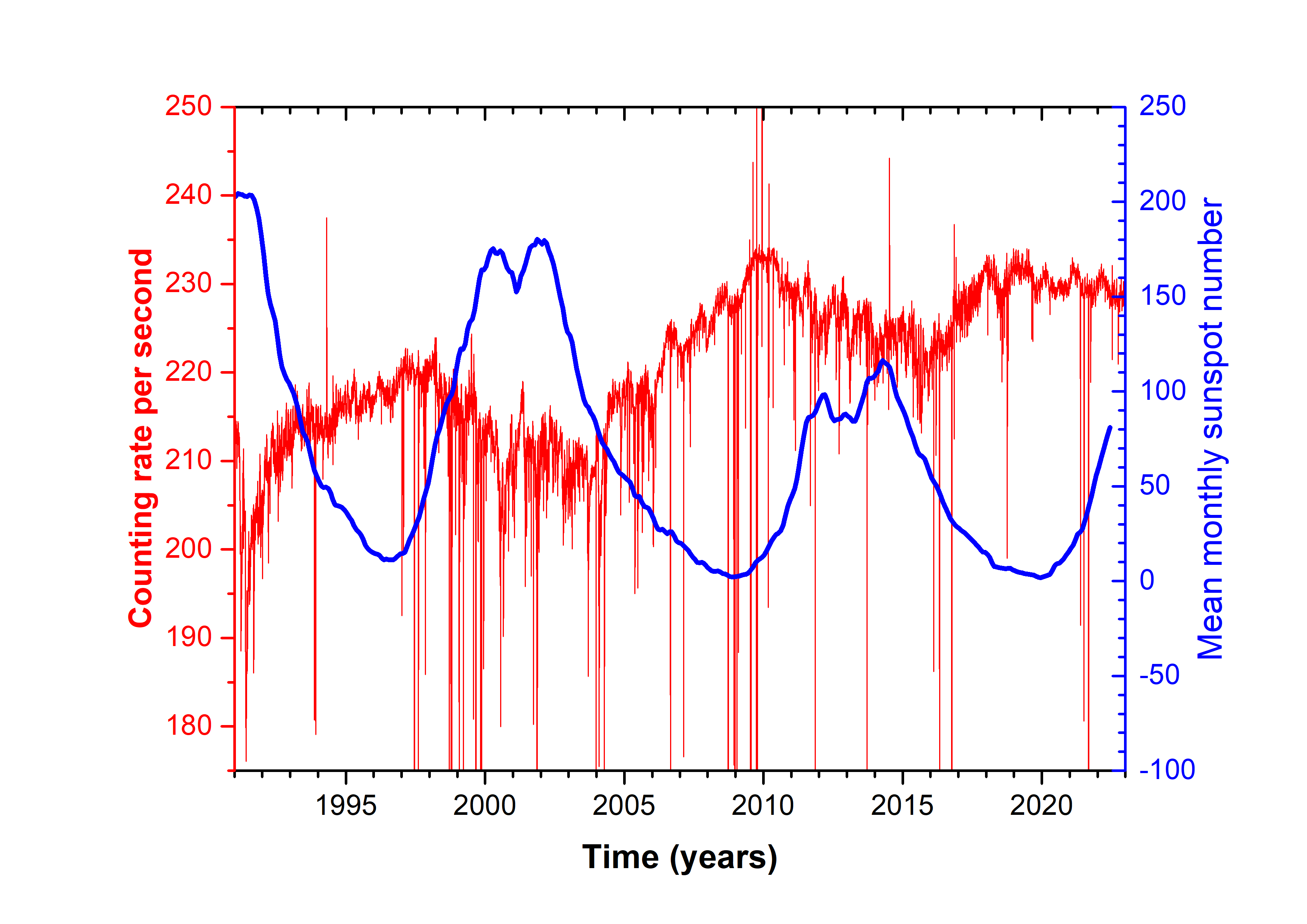}
\vspace*{-0.0cm}
\caption{Left red scale: Pressure \& efficiency corrected cosmic ray count rate profiles, from  Mexico-City NM, from 1991 to 2022. Right blue scale: monthly sunspot number profiles for the same period.
}
\label{mexico_full}
\end{figure*} 

Other measurements indicate that the amplitude of the anisotropy in the secondary cosmic ray flux due to the C-G effect is no higher than 0.1\% \citep{clay97}. On the other hand, measurements of the C-G effect in spacecraft, through the hydrogen flux in the keV energy range, indicates that the C-G effect distorts the hydrogen flux. Monte Carlo simulations show that in the ram frame where the spacecraft is toward the emission source, the C-G effect forces the hydrogen flux to the ecliptic plane, while the opposite occurs in the anti-ram frame \citep{zirn13}.
The energy spectrum of galactic cosmic rays follows a power law, from energies around 
$\sim$ 8.0 GeV to ultra-high energies. Also, the energy spectrum decreases as energy increases. In detectors whose location has a geomagnetic rigidity cutoff above 9.0 GV, 80\% of secondary particles at ground level come from galactic cosmic rays (mainly protons) of up to 2 TeV \citep{dass12}.
Places with a high cutoff in geomagnetic rigidity and located at small geographic latitudes are more effective for observing the C-G effect. However, there is a side effect, the greater the rigidity cutoff, the smaller the cosmic ray flux. Thus the observation of the C-G effect requires long periods of observation.

However, we show that some NMs, especially those located at low latitudes and with a high geomagnetic rigidity cutoff and altitude, such as the Mexico-City NM, can observe the C-G effect.
Neutron monitors detect a large variety of secondary particles, mainly nucleons produced mainly by the galactic cosmic rays which reach the Earth's atmosphere \citep{belo18,vais21}.

The organization of paper is as follows: In Section 2, we present the data analysis with a brief description of Mexico City, including a theoretical aspect on the prediction of the C-G effect, 
followed by data analysis during two 
solar minima (2008-2019) and considering
the two years together.
We present in Section 3 a preliminary analysis of the seasonality of the C-G effect. The correlation between the phases of the C-G effect and the E-W asymmetry is analyzed in section 4. Finally, in section 5, we present a summary and  conclusions of the article.

%%%%%%%%%%%%%%%%%%%%%%%%%%%%%%%%%%%%%%%%%%%%%%%%%%%%%%%

\section{Analysis}

\subsection{Mexico-City NM data}

The Neutron Monitor Database (NMDB) (\url{http://www.nmdb.eu}) provides cosmic ray data from at least 18 neutron monitors distributed 
around the world, and operated in real-time. 
From these, the Mexico-City NM is among those located at not very high geographic latitudes ($19.33^{\circ}$ N). In addition, it has a relatively high geomagnetic rigidity cutoff (8.2 GV) and a mountain altitude (2274 m asl) \citep{sten01,varg12}. These characteristics are favorable to observing the C-G effect.

The Mexico-city NM is in continuous operation since 1990. 
Fig.~\ref{mexico_full} shows the pressure \& efficiency corrected cosmic ray time profiles from these three decades. The left (red) scale is associated with the counting rate. While the scale at right (blue) is associated with the monthly sunspot number. The well-known inverse correlation between the cosmic ray counting rate and the sunspot number is evident. 

As already indicated, data from the Mexico City NM
is available from NMDB.
Despite data were corrected for pressure and efficiency, they still show some fluctuations in the counting rate

During the minimums of the solar cycles, the Sun's magnetic field pattern is like a dipole structure. Its strength weakens and provides less shielding to the galactic cosmic rays, arriving in the Earth's environment almost with an isotropic distribution. This behavior is responsible for the inverse relationship between the galactic cosmic ray intensity and the sunspot number. 
So, during the minimums of the solar cycles, the galactic cosmic ray intensity at Earth is higher.
These characteristics are essential to observe the small anisotropy due to the terrestrial motion.

%%%%%%%%%%%%%%%%%%%%%%%%%%%%%%%%%%%%%%%%%%%

In more than 30 years of continuous operation, Mexico-City NM has recorded three solar minima.
The start (minimum) of solar cycles  23, 24, and 25 was in  August 1996, December 2008, and December 2019, respectively
 \citep{hath15,pish19}.
 The maxima cosmic ray counting rate during the minima of cycles 24 and 25 was around 5\% higher than the maximum counting rate from cycle 23. Fig.~\ref{mexico_full} summarizes the situation.
 
\begin{figure*}[]
\vspace*{+0.0cm}
\hspace*{-0.00cm}
\centering
\includegraphics[clip,width=0.53
\textwidth,height=0.4\textheight,angle=0.] {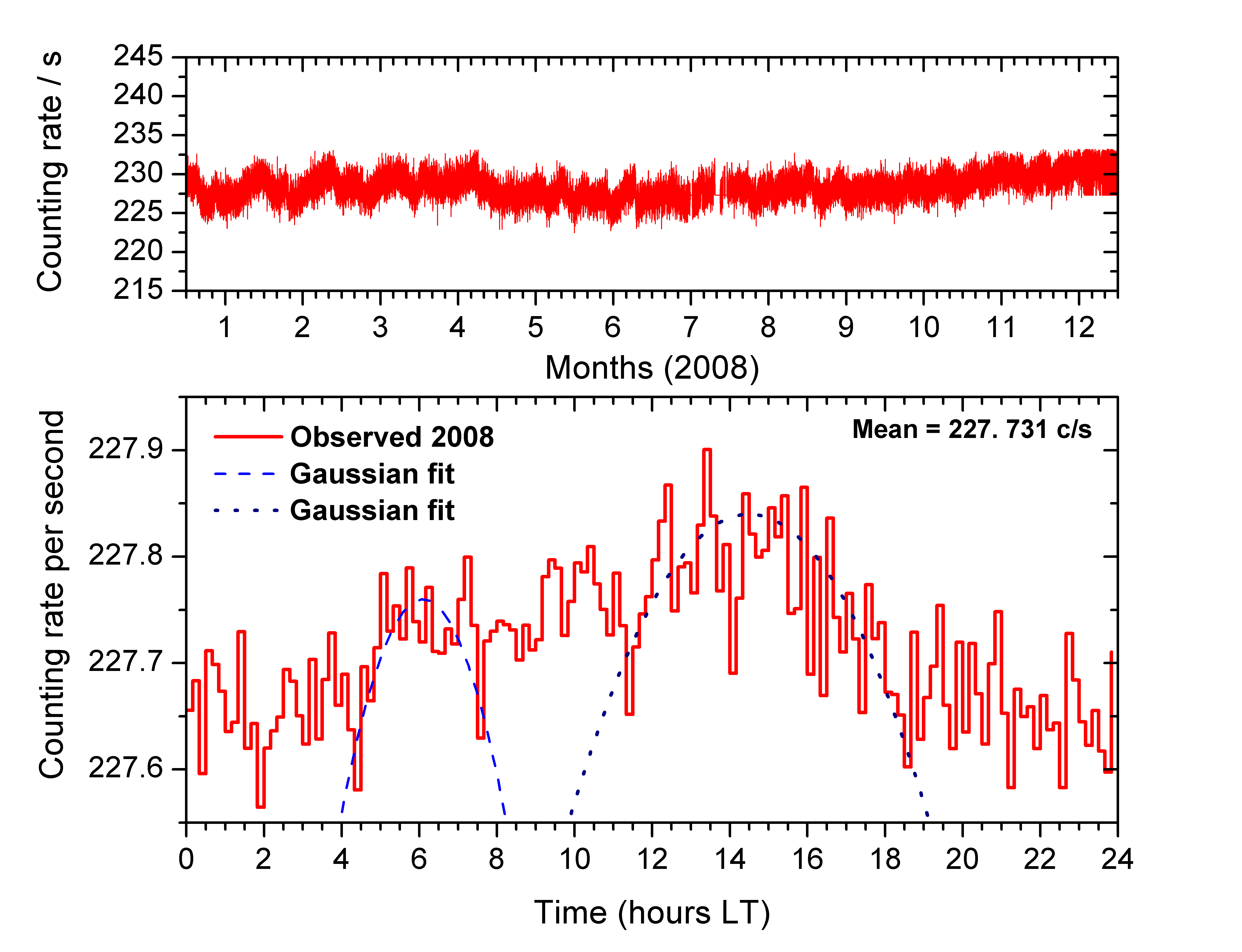}
\vspace*{-0.0cm}
\caption{Pressure \& efficiency corrected counting rate, according to Mexico-City NM data. 
Top panel: monthly (averaging 10 min) counting rate
 during 2008. Bottom panel: hourly (averaging 10 min) counting rate during 2008. The bottom panel includes the two Gaussian fits for the C-G effect and diurnal solar variation, respectively.
}
\label{mexico_gaus1}
\end{figure*}
 
\begin{figure*}[]
\vspace*{-1.0cm}
\hspace*{-0.00cm}
\centering
\includegraphics[clip,width=0.53
\textwidth,height=0.4\textheight,angle=0.] {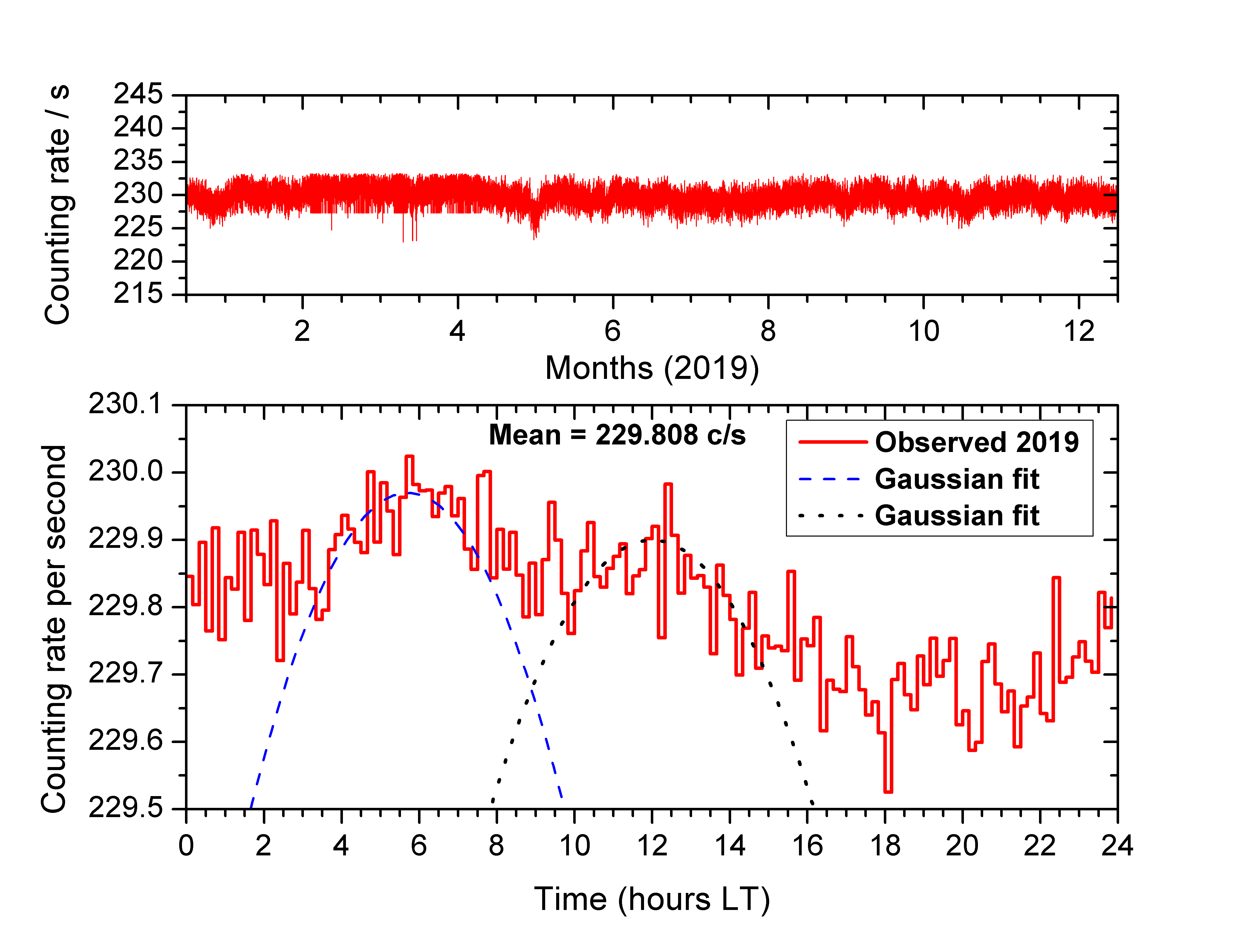}
\vspace*{-0.0cm}
\caption{Same as Fig.~\ref{mexico_gaus1}, but for the year 2019.
}
\label{mexico_gaus2}
\end{figure*} 

\begin{figure*}[]
\vspace*{-3.0cm}
\hspace*{-0.00cm}
\centering
\includegraphics[clip,width=0.8
\textwidth,height=0.4\textheight,angle=0.] {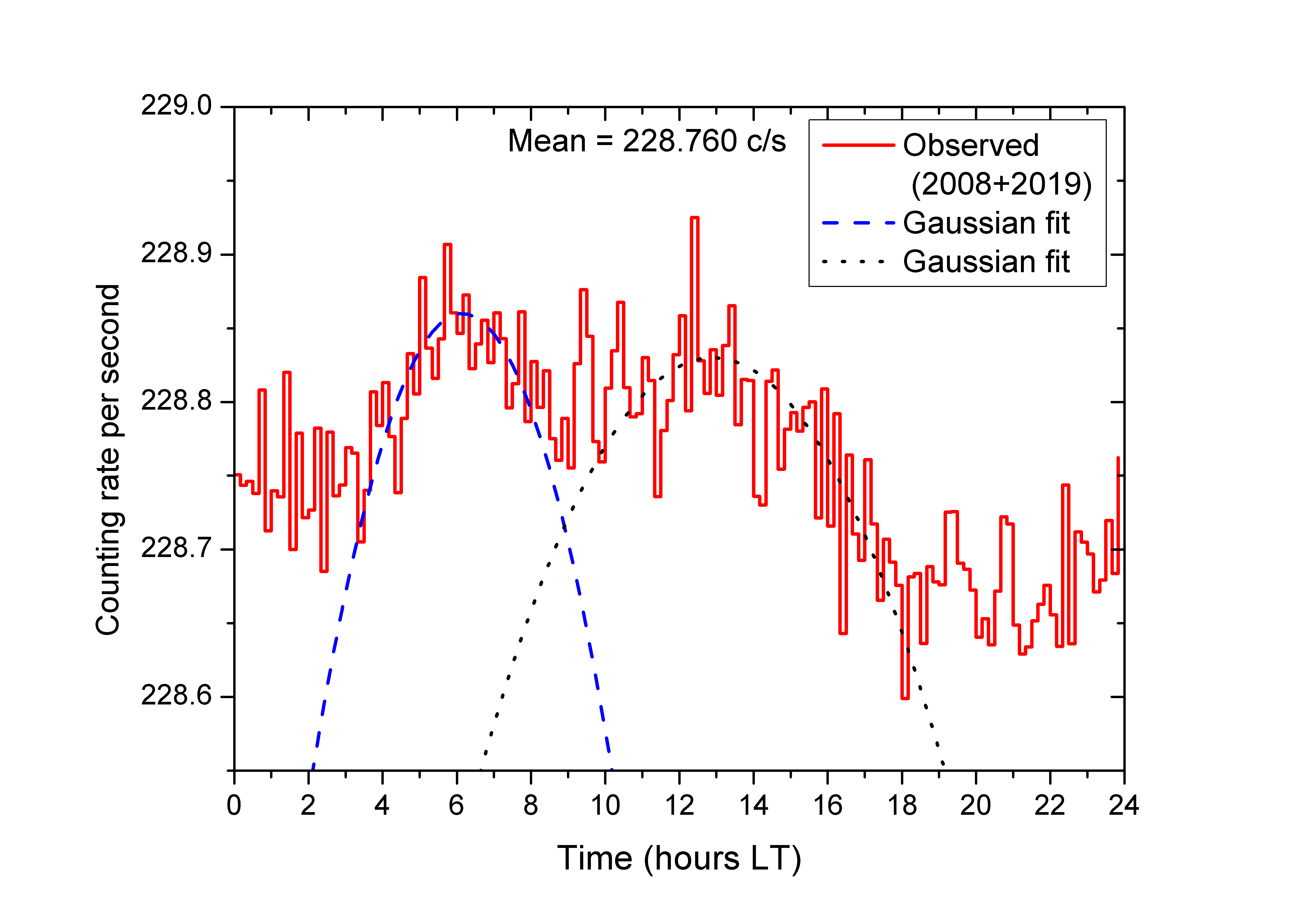}
\vspace*{-0.0cm}
\caption{Hourly (averaging 10 min) counting rate from Mexico-City NM data, 
during combined years 2008-2019 (corrected by pressure \& efficiency). 
 The figure includes the two Gaussian fits for the C-G effect and diurnal solar variation, respectively.
}
\label{mexico_gaus12}
\end{figure*} 

%%%%%%%%%%%%%%%%%%%%%%%%%%%%%%%%%
\begin{table*}[]
\vspace*{-0.0cm}
\caption{Amplitude variation and phase to the Compton-Getting (C-G) effect
and for the Solar Diurnal (SD) variation. The observation is from Mexico-City NM data.}
  \centering
  \begin{tabular}{l|crcr}
  \hline
  \hline
     & C-G Amplitude (\%) & C-G Phase LT & SD Amplitude (\%)& SD Phase LT\\
    \hline
    Observed (2008) & 0.006$\pm$ 0.006 & 06.17$\pm$1.20 &  0.033$\pm$ 0.066  &  14.57$\pm$2.04\\
    Observed (2019) & 0.073$\pm$0.066 & 05.70$\pm$1.36 & 
    0.042$\pm$0.040 & 12.00 $\pm$ 1.54 \\
    Obs.(2008-2019) & 0.043$\pm$ 0.019& 6.15$\pm$ 1.71 & 0.031$\pm$ 0.014  & 12.90$\pm$ 1.97      \\
    Predicted & 0.044        & 06:00 &  $<$0.6          &  15:00-18:00\\ 
    \hline
    \hline
    \end{tabular}
    \label{tab:1}
\end{table*}

\subsection{Compton-Getting effect prediction}

Assuming that an isotropic galactic cosmic rays flow reach the Earth with a power-law energy spectrum, such as $E^{-\gamma}$, with
$\gamma = 2.7$, a ground-level detector would see an increase in the count rate when its field of view looks along the direction of Earth's orbital motion, predicted as
 \citep{glee68}
\begin{equation}
\frac{\Delta f}{f}=(\gamma+2) \frac{v}{c} \cos \lambda,
\end{equation}
where $v$ is the mean speed of Earth's orbital motion (29.78 km/s). Here, we assume that the galactic cosmic rays speed is close to c (the speed of light in a vacuum). This approximation is valid
for cosmic rays with energies greater than GeV, and 
$\lambda$ is the angle between the direction of the detector's sensitivity and Earth's velocity vector. For non-directional (large field of view) detectors such as NMs, $\lambda$ is close to the geographic latitude angle where the detector is located.
The C-G effect amplitude variation predicted  to the Mexico-City NM is
0.044\% and a phase at 06:00 LT.

\subsection{Cosmic ray anisotropies at Earth}

In the region of energies below $\sim$50 TeV, the observed anisotropies of galactic cosmic rays mean that their propagation in the inner heliosphere (interplanetary space) is not isotropic. There are gradients parallel and perpendicular to the ecliptic plane. They are responsible for diurnal solar anisotropy and North-South anisotropy, among other smaller ones. Cosmic ray count rate at ground level provides valuable insight into the processes described above \citep{asip09}.

The large anisotropy observed at ground level is the diurnal solar anisotropy. Galactic cosmic ray propagation (with hardness below $\sim$50 GV) in the plane of the ecliptic corotates with the interplanetary magnetic field (IMF) \citep{park64,axfo65}. When the flux reaches Earth, it produces the diurnal solar anisotropy, approximately perpendicular to the Sun-Earth line, with a phase about 15:00 LT (under positive solar cycle polarity $qA>0$) and about 18:00  LT (under negative solar cycle polarity $qA<0$) \citep{sabb13}. Like the C-G effect, the diurnal anisotropy depends on geographic latitude $\lambda$ as $\cos \lambda$ \citep{rao72}. Thus, the greater the latitude of the place, the smaller the amplitude of diurnal solar anisotropy, and it disappears in the polar regions.

Here we highlight an anisotropy due to the Earth's orbital motion, the so-called C-G effect, through analysing the daily counting rates at Mexico City NM. We select three years, 1996, 2008, and 2019, during the solar minima when the galactic cosmic ray flux reaching Earth is maximum. In these years, the Sun's magnetic field is weakened. It is like a well-behaved magnetic dipole. Consequently, the Sun's magnetic effects on cosmic rays are smaller.

However, 1996 data from Mexico-NM are not available
at NMDB at 1 h or fewer time intervals, which
impossibilities the analysis. So, 1996 NM data were not include. The analysis is restricted only to 2008 and 2019 years.

Let's start with the 2008 data (minimum start of cycle 24). Fig.~\ref{mexico_gaus1} (top panel) shows the monthly counting rate time profiles (averathing 10 min) from Mexico-City NM data. 
Already Fig.~\ref{mexico_gaus1} (bottom panel) shows the mean count daily rates for the full year 2008. The two curves are two Gaussian fits around the first and second peaks. The first peak is identified as due to the C-G effect, originating by the terrestrial motion, and the second one as the diurnal solar variation due to solar wind corotating structure reaching the Earth. 

Table~\ref{tab:1} indicates the amplitude variation in percentage relative to the daily mean counting rate in 2008 due to the C-G effect and diurnal solar variations, respectively. Table~\ref{tab:1} also 
indicates the phase in LT for these two peaks. 

Note that in 2008 the amplitude variation due to daily variation is around 18\% greater than the amplitude variation due to the C-G effect.

Fig.~\ref{mexico_gaus2} shows the same type of analysis for the 2019 data (minimum start of cycle 25). Also, Table~\ref{tab:1} indicates the amplitude variation in percentage relative to the daily mean counting rate and the phase in LT due to the C-G effect and diurnal solar variations, respectively.

In most cases, the daily solar variation amplitude is higher than the C-G effect amplitude. However, note that in 2019, there is an anomaly. The amplitude variation due to the C-G effect is around 57.5\% greater than the amplitude variation due to diurnal solar variation.

Finally, Fig.~\ref{mexico_gaus12} represents a combined analysis including data from the 2008 and 2019 values. Again the 
Table~\ref{tab:1} shows the values found for the amplitude variations due to the C-G effect and diurnal solar variation, respectively, as well as their phases. 

We would point out that the amplitude variation and phase for the C-G effect obtained in the combined data analysis (2008-2019) agree with expected values.

\begin{figure*}[]
\vspace*{+0.0cm}
\hspace*{-0.00cm}
\centering
\includegraphics[clip,width=0.65
\textwidth,height=0.4\textheight,angle=0.] {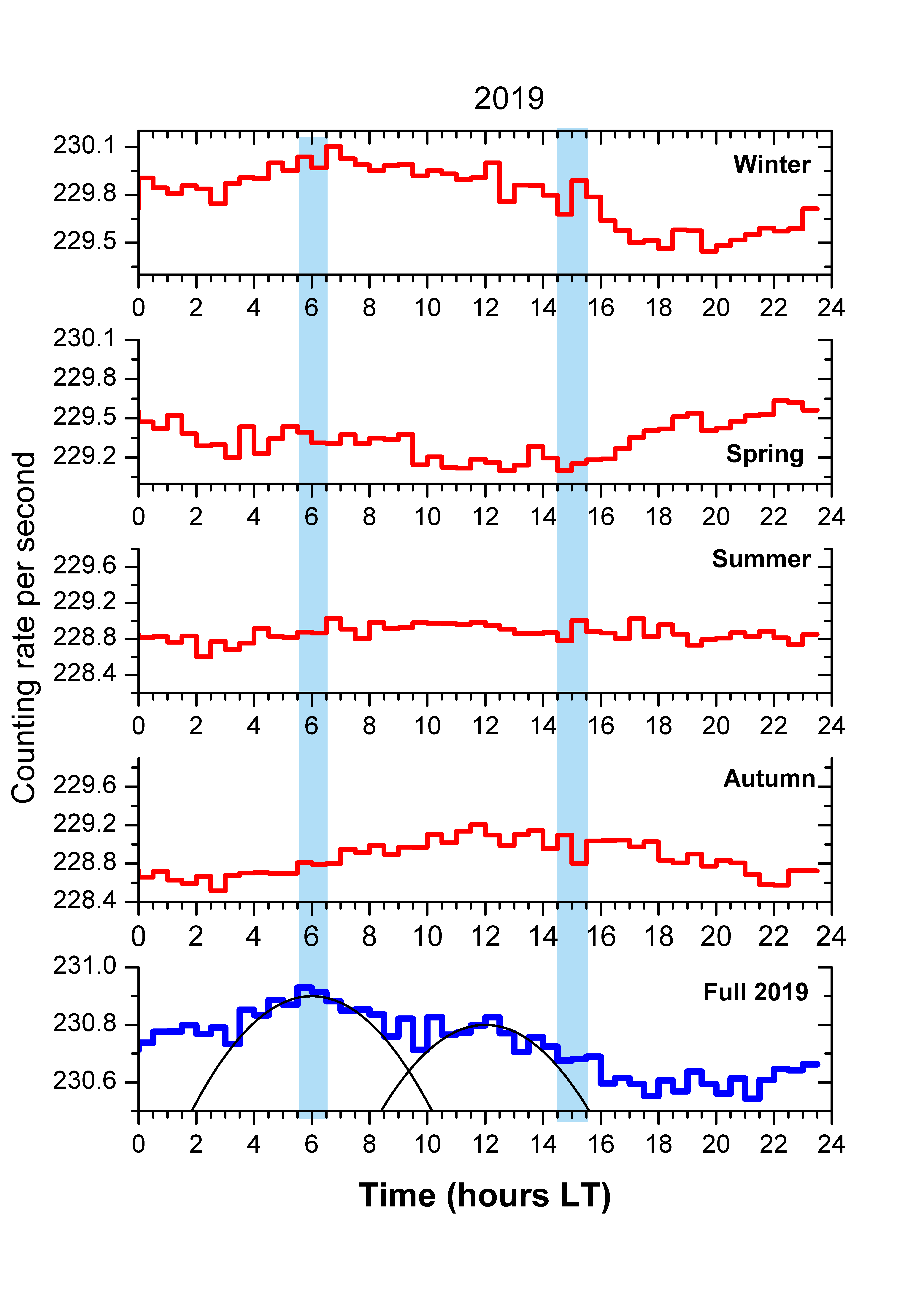}
\vspace*{-0.0cm}
\caption{Pression \& efficiency corrected counting rate, according to Mexico-City NM data. 
Top panel: montly (averaging 10 min) counting rate
 during 2008. Bottom panel: hourly (averaging 10 min) counting rate during 2008. The bottom panel includes the two Gaussian fits for the C-G effect and diurnal solar variation, respectively.
}
\label{seasonal2}
\end{figure*}

\section{Seasonality of the C-G effect}

Despite low statistics, we also examine the cosmic-ray diurnal variation from Mexico-City NM data over 2019, according to the seasons.
Dates regarding season here used are 
Winter (01 January to 20 March); 
Spring (20 March to 21 June);
Summer (21 June to 23 September);
Autumn (23 September to 31 December).

We see a significant variation, as shown in Fig.~\ref{seasonal2}, showing the hourly (averaging 30 min) counting rate per second during the 12 (seasonably) months of 2019.

From Fig.~\ref{seasonal2}, we also can see that the count rate during winter is the one that most contributes to the C-G effect observation. 
We also can observe that the counting rate during autumn does not contribute to the C-G effect observation but contributes significantly to diurnal solar variation.

Fluctuation on the interplanetary magnetic filed and the E-W asymmetry of the cosmic ray flow (see next section) are responsible for the phase fluctuations of the Compton-Getting effect and of the diurnal solar variation around values, in most cases, earlier than expected. The details of this mechanism are the subject of the next section.

%%%%%%%%%%%%%%%%%%%%%%%%%%%%%%%%%%%%%%%%%%
\section{C-G effect and the West-East asymmetry}

From the 1930s onwards, it's known that 
the flow of cosmic rays reaching Earth from the West is higher than
from the East; and is known as the East-West (E-W) effect
\citep{kami63,dorm67}. This effect is due to the de
deflection of the cosmic
ray-charged particles by the Earth's magnetic field.
The count excess from the West direction means that
primary cosmic rays are predominantly positively charged
particles.

The observation of muons data detected by underground telescopes, allows to obtain the E-W asymmetry of their parent particles, galactic cosmic rays with high magnetic rigidity (above 1000 GV) \citep{yasu91}. 

So, cosmic ray directional telescopes can obtain the phase from the E-W asymmetry, and from it obtain the phase of the C-G effect adding (clockwise) a right angle ($=+6$ hr) to E-W asymmetry phase:
\begin{equation}
Phase(C-G) = Phase(E-W) + 6\; hr.
%\label{eq_phase}
\label{phases}
\end{equation}

\begin{figure}[]
\vspace*{+0.0cm}
\hspace*{-0.00cm}
\centering
\includegraphics[clip,width=0.34
\textwidth,height=0.4\textheight,angle=0.] {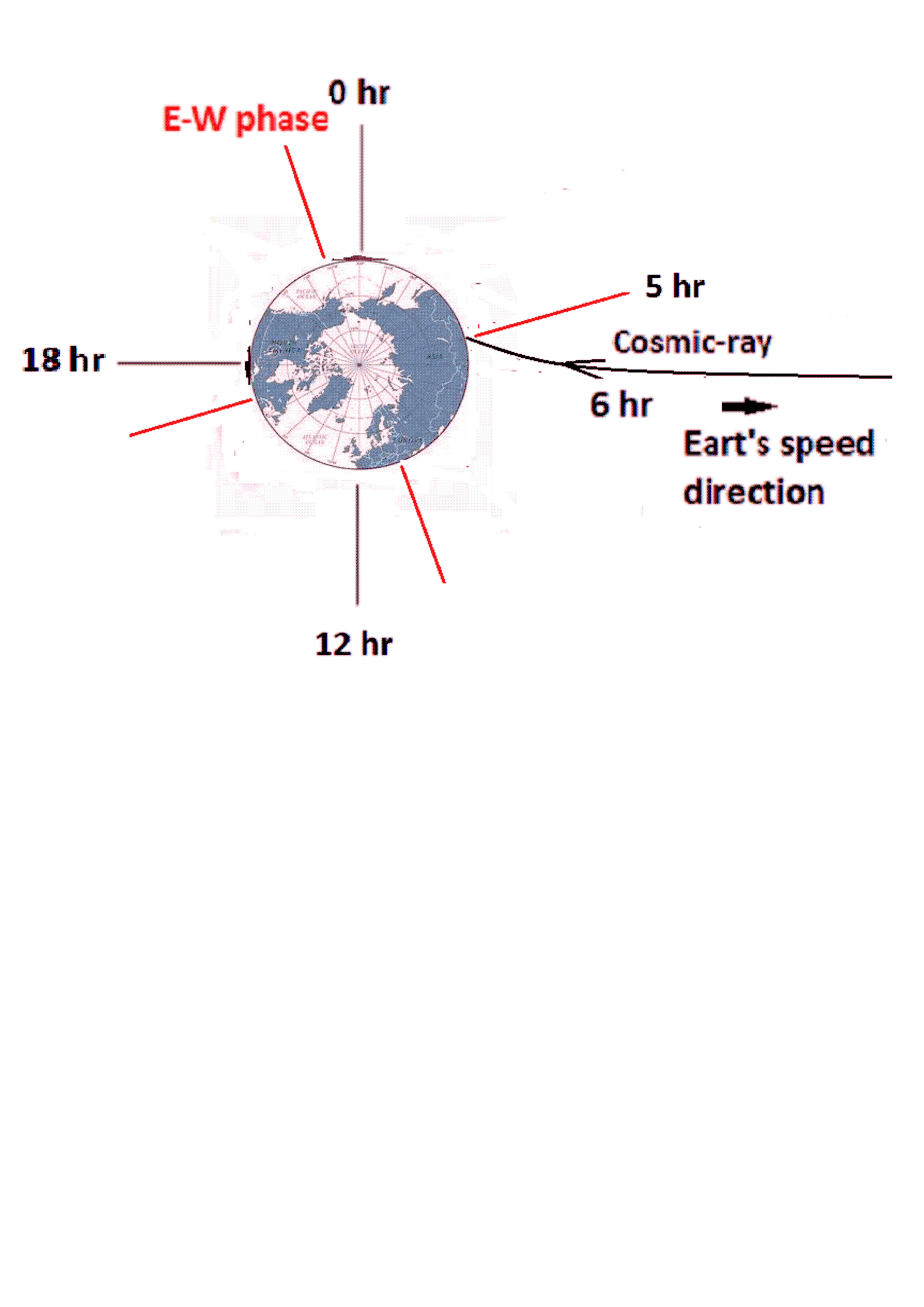}
\vspace*{-4.0cm}
\caption{Scheme showing how to obtain the East-West asymmetry phase from the C-G effect phase or vice versa.
Deflection of cosmic rays by Earth's magnetic field
  in the west direction is responsible for the phase
of the C-G effect be about 1 hour earlier than expected.
}
\label{mexico_vector}
\end{figure}

The E-W asymmetry is responsible for changing the phase of the C-G effect to times earlier than expected. Fig.~\ref{mexico_vector} details this mechanism.

The E-W asymmetry can also explain the shift of the phase of the diurnal solar variation for early times than expected. In the case of 2019 (beginning of solar cycle 25) with the solar cycle's  polarity conditions ($qA>0$), the expected phase for the diurnal solar variation is 15 hr LT, as shown by the right arrow
at the top of Fig.~\ref{seasonal2} where shift values of up to 3 hr relative to the expected value is observed, while the average value in the (averaging 10 m) data is  12.90$\pm$ 1.97 LT (see 
Table~\ref{tab:1}).

%%%%%%%%%%%%%%%%%%%%%%%%%%%%%%%%%%%%%%%%%%%

\section{Discussions and Conclusions}

We present a study on the daily variation of count rate at the Mexico City NM detector located at low geographic latitude ($19.33^{\circ}$N) and a high geomagnetic rigidity cutoff (8.2 GV). We highlight the study of anisotropy due to the terrestrial orbital movement known as the C-G effect.

To optimize the cosmic ray detection at ground level (2274 m asl), we used the 2008 and 2019 data, which coincides with the last two solar minima, when the flow of galactic cosmic rays reaching Earth is higher, and the Sun's magnetic field is weakened,
with a dipolar pattern, and their influence on galactic cosmic rays is lowest.

Two modulations in the cosmic ray flux still survive in the Mexico-NM detector, even after corrections for variations in barometric pressure and efficiency. The first is in the early hours of the day (expect 6:00 hr LT), consistent with a modulation due to the Earth's orbital motion, i.e., the C-G effect. predicted by \cite{comp35} as a first-order relativistic effect.

The second one is the known diurnal solar variation, in the early hours
 of the afternoon (expected between 15:00 and 18:00 LT),
The cosmic ray diffusion theory \citep{park64,axfo65}, including
other factors, such as the scattering effect due to irregularities in the solar magnetic field \citep{joki69,levy76} and latitude effects \citep{rao72},
can describe the diurnal solar variation.

Table~\ref{tab:1} summarizes the results, including the counting rate variation and the phase of both anisotropies, the C-G effect, and diurnal solar variation, respectively. Both the counting rate variation and phase for the C-G effect, obtained from combined data (2008-2019) from Mexico City NM, agree with those predicted.

We have also looked for the seasonality of the C-G effect, with the seasons during 2019. We observe a significant variation. The same is true for the solar diurnal variation. However, due to the low statistics, the results on the seasonality of the two anisotropies are preliminary. 

Even so, the data obtained during the winter are the ones that contribute the most to the C-G effect. The data collected in autumn has no contribution to the C-G effect but significantly contributes to diurnal solar variation. Fig.~\ref{seasonal2} summarizes the results.

Finally, we show that in addition to fluctuations in the interplanetary magnetic field, the E-W effect on the cosmic ray flow reaching Earth is also responsible for 
dispersion, observed in the phases of both anisotropies. In most cases, the E-W effect is responsible for bringing the phases forward, sometimes several hours ahead of the expected ones. In this study, this effect is more pronounced in diurnal solar variation. We highlight a correlation between the phases of the E-W and C-G effects (see Eq.~/ref{phases})\citep{yasu91}. In the present case,
the C-G effect phase for the combined data (2008-2019) is 
06.15 LT. Consequently, the E-W phase will be close to 0.15  LT. This result is a prediction for the E-W phase at the Mexico City location.

%%%%%%%%%%%%%%%%%%%%%%%%%%%%%%%%%%%%%%%%

\section{Acknowledgments}

We acknowledge the NMDB database (www.nmdb.eu), founded under the European Union's FP7 programme (contract no. 213007) for providing data.
Mexico City neutron monitor data were kindly provided by the Cosmic Ray Group, Geophysical Institute, National Autonomous University of Mexico (UNAM), Mexico.
This work is supported by Fundacao de amparo a Pesquisa do Estado do Rio de Janeiro (FAPERJ)  under Grant E-26/010.101128/2018.

\newpage

\bibliography{lucyC}{}
\bibliographystyle{aasjournal}

\end{document}